\documentstyle[11pt,aaspp4]{article}

\received{1998 March 27}
\accepted{1998 June 10}

\journalid{507}{1998 November 1}
\articleid{16}{23}

\slugcomment{POPe-755, UTAP-286, RESCEU-16/98}

\lefthead{Suginohara \& Ostriker}
\righthead{Density profile of hot gas in clusters of galaxies}

\begin{document}
\title{The Effect of Cooling 
on the Density Profile of Hot Gas in Clusters of 
Galaxies: Is Additional Physics Needed?}
\author{Tatsushi Suginohara\altaffilmark{1,2,3}
and Jeremiah P. Ostriker\altaffilmark{4}}
\affil{Department of Astrophysical Sciences, Princeton University,
Princeton, NJ 08544}

\altaffiltext{1}{Department of Physics, the University of Tokyo,
Tokyo 113, Japan}
\altaffiltext{2}{RESCEU, School of Sciences, the University of Tokyo,
Tokyo 113, Japan}
\altaffiltext{3}{tatsushi@astro.Princeton.EDU, JSPS Postdoctoral Fellow}
\altaffiltext{4}{jpo@astro.Princeton.EDU}
\begin{abstract}
We use high-resolution hydrodynamic simulations
to investigate the density profile of hot gas in clusters of galaxies,
adopting a variant of cold dark matter cosmologies and
employing a cosmological N-body/smoothed particle hydrodynamics code to
follow the evolution of dark matter and gas.
In addition to gravitational interactions, gas pressure, and shock
heating, we include bremsstrahlung cooling in the computation.
Dynamical time, two-body relaxation time, and cooling time
in the simulations are examined to
demonstrate that the results are free from artificial relaxation
effects and that the time step is short enough to accurately follow
the evolution of the system.
In the simulation with nominal resolution of $66h^{-1}\,{\rm kpc}$
the computed cluster appears normal, but
in a higher (by a factor 2) resolution run,
cooling is so efficient that the final gas density profile shows
a steep rise toward the cluster center
that is not observed in real clusters.
Also, the X-ray luminosity of $7\times10^{45}{\rm ergs\,s^{-1}}$
far exceeds that for any cluster of the computed temperature.
The most reasonable explanation for this discrepancy is that 
there are some physical processes still missing in the
simulations that actually mitigate the cooling effect and
play a crucial role
in the thermal and dynamical evolution of the gas near the center.
Among the promising candidate processes are heat conduction and heat input
from supernovae.
We discuss the extent to which these processes can alter the
evolution of gas.
\end{abstract}

\begin{center}
{\it The Astrophysical Journal}, 507, 16 (1998)
\end{center}

%
%
\clearpage
\section{Introduction}

Hot X-ray--emitting gas in clusters of galaxies 
contains a variety of information relevant to many fields
in astrophysics.
Density and temperature profiles of the gas give
among the most reliable estimates of cluster mass,
which are of unparalleled importance to cosmology
(e.g., \cite{bah95}).
They will also reflect physical processes that have played
a crucial role in the thermal and dynamical evolution of the gas.
In addition to gravitation, hydrodynamics, and shock heating,
such processes may include radiative cooling, heat conduction,
and feedback from star formation.

One of the central questions about the cluster gas structure
involves the presence of a core.
Observations have revealed that the gas density profile has a
distinct core, inside of which density is nearly constant.
Gravitational N-body simulations have demonstrated that
gravity alone cannot produce such a core in an object
formed as a result of hierarchical structure formation;
halos formed in high-resolution N-body simulations 
have density profiles with significant slope toward the center
up to the resolution limit
(\cite{nav96}, 1997; \cite{fuk97}; \cite{moo98}).
For example, Navarro, Frenk, and White (1996, 1997) found that
the density profiles of halos with a wide range of masses can be
fitted by
\begin{equation}
\label{eqn:nfw}
\rho(r) = \frac{\rho_c}{(r/r_s)(1+r/r_s)^2}
\end{equation}
(hereafter referred to as the NFW profile), with
$\rho_c$ and $r_s$ being the fitting parameters.
In the NFW profile, gas temperature would approach zero in a
cluster's central regions ($T \propto r$) were it to have the same
profile as dark matter.
While this model does produce a convergent X-ray luminosity,
the temperature structure is different from what is observed.
It follows that processes other than gravity are responsible for
the formation of the cores.
Makino, Sasaki and Suto (1998) pointed out that
the gas distribution develops a core in a dark matter halo
having a central cusp (e.g., the NFW profile)
if the gas is isothermal and in hydrostatic
equilibrium.
Even if the gas obeys a less stringent limit of constant entropy
($T/n^{2/3} \to {\rm const}$), 
it would avoid a central cusp and have an apparently 
constant-density, constant-temperature core.
But it is not evident what physical mechanisms could enforce
either isothermality or isentropy.

Many authors have studied the evolution of clusters using simulations
that include hydrodynamics
(\cite{evr90}; \cite{tho92}; \cite{kat93}; \cite{bry94a}, 1994b; 
\cite{kan94}; \cite{nav95}; \cite{bar96}; \cite{bry97}, 1998;
\cite{eke98}; \cite{pen98}; \cite{yos98}).
In general these simulations have succeeded in producing clusters that 
have cores similar to those observed.
However, the situation is far from satisfactory 
for at least two reasons.
First, artificial two-body relaxation may affect the dynamics,
especially in the central region.
Indeed, Steinmetz and White (1997) showed that this effect gives rise to
artificial energy transfer from dark matter to gas.
Both spatial and mass resolutions are only marginally adequate
in most published works.
Second, almost none of the simulations include radiative cooling
of gas.
Although cooling is probably unimportant in the outer part of a
cluster (\cite{sar86}), it may affect the dynamics in the central region.

This work is an investigation of cluster gas using hydrodynamic
simulations that attempt to address these limitations.
First, we 
minimize two-body relaxation effects for a given computational cost
by employing a multiresolution technique
that enables us to improve resolution only inside the clusters
where we really need high resolution.
Second, we include cooling due to bremsstrahlung,
which dominates cooling of gas at above $\gtrsim 10^7{\rm K}$.
We ignore line cooling, but this is not a serious problem
because we focus our attention on X-ray--emitting gas.
A practical reason for ignoring line cooling is that doing so enables us to
avoid very short timescales in moderate temperature ($10^4$ -- $10^6
{\rm K}$), high density regions.
Allowing for line cooling would strengthen the conclusions of
this paper.
We organize the rest of the paper as follows.
In \S\ 2 we describe the method and parameters of the
simulations.
We present the results in \S\ 3.
As we will see, the most important result is that 
cooling and increased resolution give rise to
a density profile of the gas that rises steeply toward the center
and consequently produces excessive X-ray luminosity.  In \S\ 4
we discuss the implications of our results in
connection with physical processes that are still missing in the simulations.
In \S\ 5 we give our conclusions.

%
%
\section{Simulations}

%
We work with a cold dark matter--dominated universe with $h = 0.6$,
$\Omega_0 = 1$, $\Lambda=0$, $\Omega_b = 0.07$,
$n_{\rm spect}=0.81$, and $\sigma_8 = 0.50$, where $h$ is the 
Hubble constant 
in units of $100\,{\rm km\,s^{-1}Mpc^{-1}}$, $\Omega_0$ is
the cosmological density parameter, $\Lambda$ is the 
cosmological constant, $\Omega_b$ is the baryon density parameter,
$n_{\rm spect}$ is the power-law index of the primordial 
density fluctuations,
and $\sigma_8$ is the rms density fluctuations on 
$8\,h^{-1}{\rm Mpc}$ scale.
These parameters are chosen so that the universe is at least
marginally consistent with all of the following observations:
the amplitude of cosmic microwave background anisotropies
observed by COBE,
the observational bound on the Hubble constant, 
the abundance of
light elements, the baryon fraction in clusters,
the number density of clusters, and the age of the globular clusters.
The model is similar to the tilted cold dark matter (TCDM)
model of Cen and Ostriker (1993).
We employ a cosmological N-body + smoothed particle
hydrodynamics (SPH) code (see e.g. Monaghan 1992 for a review
of SPH).
The N-body part of the code uses the Barnes-Hut (1986)
tree algorithm and is described in
Suginohara et al. (1991).
The SPH part of the code is described in Suginohara (1992).
Overall, the code is similar to the one described in
Katz, Weinberg, and Hernquist (1996).
An application of the earlier version of the code is in Suginohara
(1994, 1995).

We carry out a multiresolution simulation that allows us to achieve
high resolution only inside the clusters.
We prepare two sets of initial conditions
with different mass resolution
in a cubic box with comoving size $L = 42\,h^{-1}{\rm Mpc}$.
These sets have exactly the same realization,
up to the Nyquist wavenumber corresponding to the lower resolution,
of the theoretical power spectrum of density fluctuations at $z=19$.
The number of particles in the higher resolution realization is
8 times that in the lower resolution one.
In each realization we assign to all particles
the same gravitational softening length $\varepsilon$, which is
fixed in the comoving coordinates and is
equal to 0.1 times the mean interparticle separation of
dark matter particles.
The initial gas temperature is set to $10^4\,{\rm K}$.
Then we use the following two-step procedure.
First, we evolve the system from the lower resolution realization.
When the low-resolution (LR) run
has been carried out up to the present epoch
we identify cluster centers.
Then we come back to the initial distribution of particles
and identify those regions that will end up with overdensity
larger than 140 around the cluster centers.
We replace the particles in the identified regions with those
in the high-resolution realization.
Then, starting from this initial condition,
we run the second simulation,
which is hereafter referred to as the multiresolution (MR) run.
Katz and White (1993) and Navarro, Frenk, and White (1996, 1997)
used a similar method to construct initial conditions with
variable resolution.

The ratio of the number of gas particles to that of dark matter
particles is always set as 1/8.
Thus in either the low-resolution or high-resolution realizations,
the masses of gas particles, $m_g$,
and dark matter particles, $m_d$, are comparable
($m_g/m_d = 0.6$).
Since two-body relaxation time
is inversely proportional to $\max(m_g, m_d)$,
it is maximized when $m_g = m_d$
under the condition that the available CPU time is fixed.
Thus, with the available CPU cost,
the artificial two-body effect is significantly reduced
compared to when we used an equal number of gas and
dark matter particles, as has been done in most previous work.
Indeed, the mass of a dark matter particle for the high-resolution 
realization, $1.5\times10^{10} M_\odot$,
is less than one-third of the upper bound 
for two-body heating to be subdominant
(\cite{ste97}) for the cluster
that is mainly discussed in the following sections.

We allow each particle to have its own time step.
The time step of each particle satisfies $\Delta t = 
(\Delta t)_{\rm sys} / 2^{n_{\rm step}}$, 
where $(\Delta t)_{\rm sys} = 2.4 \times 10^6 \,{\rm yr}$.
The time-step index $n_{\rm step}$, which is a nonnegative
integer, is chosen so that the following three conditions
are satisfied: 
(i) the Courant condition is satisfied with the Courant number taken
to be 0.3 (\cite{kat96}),
(ii) expected relative change
in the internal energy within a time step does not exceed 0.05, and 
(iii) expected change in position within a time step does not
exceed 0.05$\varepsilon$.
The parameters for the simulations are summarized in Table~
\ref{tab:param}.

Bremsstrahlung cooling is included in the simulations.
Throughout this paper
we assume the mass fraction of helium, $Y$, to be 0.24
and the effect of metals to be negligible,
and we use the approximation that
$\bar g = 1.2$,
where $\bar g$ is the frequency averaged Gaunt factor.
Then bremsstrahlung emissivity is given by
\begin{equation}
\label{eqn:brems}
   \Lambda_B = 5.2\times10^{-28} T^{1/2}n^2\,{\rm ergs\,s^{-1}\,cm^{-3}},
\end{equation}
where the gas temperature $T$ is in units of K
and the gas density $n$ is in units of ${\rm cm^{-3}}$
(\cite{ryb79}).

\begin{table}[t]
\caption{Simulation parameters}
\label{tab:param}
\begin{center}
\begin{tabular}[t]{ccrrcccc}
   \hline
   Run & Resolution $^{\rm a}$ &
   $N_g{}^{\rm b}$  & $N_d{}^{\rm c}$ &
   $m_g (M_\odot) {}^{\rm d}$  &  $m_d (M_\odot) {}^{\rm e}$  &  
   $\varepsilon (h^{-1}{\rm kpc}) {}^{\rm f}$ &
   $h_{\rm min} (h^{-1}{\rm kpc}) {}^{\rm g}$ \\
   \hline
   LR  & low        &
   32{,}768         & 262{,}144       &
   $7.3\times10^{10}$          & $1.2\times10^{11}$           &
   66                                         &
   66                                         \\
   MR  & low        & 
   27{,}962         & 223{,}696       &
   $7.3\times10^{10}$          & $1.2\times10^{11}$           &
   66                                         &
   66                                         \\
   MR  & high       &
   38{,}448         & 307{,}584       &
   $9.2\times10^9$             & $1.5\times10^{10}$           &
   33                                         &
   33                                         \\
   \hline
\end{tabular}
\end{center}
\bigskip
\renewcommand{\arraystretch}{0}
\begin{tabular}{r@{}p{16cm}}
   $^{\rm a}$ & In run LR, the entire volume is represented by
                low-resolution particles.  In run MR, those
                regions which will end up in clusters are represented
                by high-resolution particles, and the remainder of
                the volume by low-resolution particles.\\
   $^{\rm b}$ & The number of gas particles.\\
   $^{\rm c}$ & The number of dark matter particles.\\
   $^{\rm d}$ & The mass of a gas particle.\\
   $^{\rm e}$ & The mass of a dark matter particle.\\
   $^{\rm f}$ & Gravitational softening length.\\
   $^{\rm g}$ & Minimum smoothing length of gas particles.\\
\end{tabular}

\bigskip
\bigskip

\end{table}
%

%
%
\section{Results}
\label{sect:results}
In this section we present results of the simulations.
We have obtained 12 clusters with $M_A \geq 2\times10^{14} M_\odot$,
where $M_A$ is the total mass inside a radius of $1.5h^{-1}{\rm Mpc}$.
We have identified these clusters in the following way.
First, we extract gas particles with gas density larger than
$10^{-4} {\rm cm^{-3}}$ out of the whole simulation.
Next we group these gas particles using the conventional 
friends-of-friends algorithm, connecting particles with separation
smaller than 0.2 times the mean interparticle separation.
We find the center of mass of each group.
Then we come back to the entire simulation and extract a spherical
region of a radius $2.5h^{-1}{\rm Mpc}$,
whose center matches the center of mass of the identified group.
Finally we redefine the cluster center as the position of the gas
particle that has the largest X-ray emissivity.

\subsection{Density and temperature profiles}
\label{subsect:profile}
\begin{figure}
\plotone{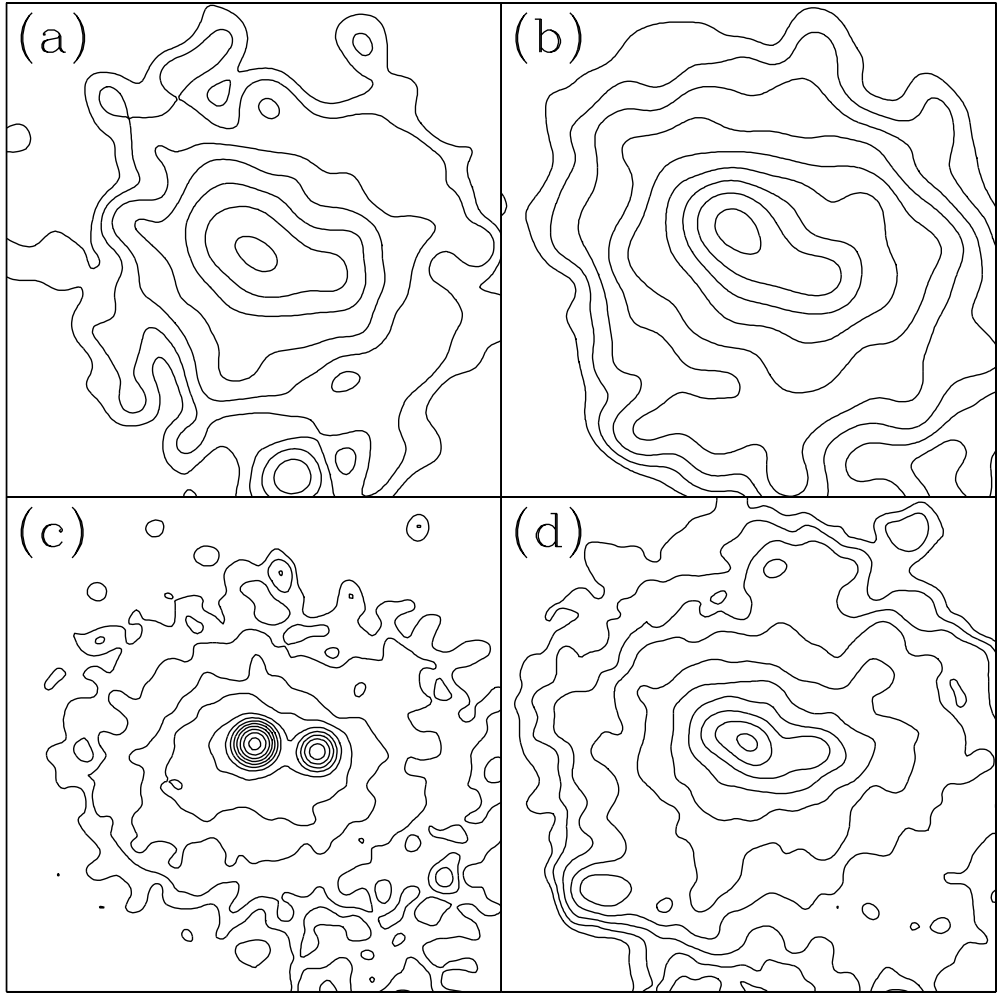}
\caption{Contours of the column density $\Sigma$ of the
richest cluster in the simulation, projected on a plane.
The neighboring
contour lines correspond to increments of 0.2 in $\log\Sigma$.
(a) Column density of the gas in run LR.  
The contour line that is nearest to the cluster center
corresponds to $\Sigma=10^{-2.6}\,{\rm g\,cm^{-2}}$.
(b) Column density of the dark matter in run LR.
The contour line nearest to the cluster center
corresponds to $\Sigma=10^{-1.4}\,{\rm g\,cm^{-2}}$.
(c) Column density of the gas in run MR.
The contour line nearest to the cluster center
corresponds to $\Sigma=10^{-1.6}\,{\rm g\,cm^{-2}}$.
(d) Column density of the dark matter in run MR.
The contour line nearest to the cluster center
corresponds to $\Sigma=10^{-1.2}\,{\rm g\,cm^{-2}}$.
\label{fig:contour}}
\end{figure}

In the remainder of the paper we concentrate,
unless otherwise stated, on
the richest cluster, with $M_A = 6.9\times10^{14} M_\odot$.
The half-mass velocity dispersion is $790\,{\rm km\,s^{-1}}$,
and the cluster is in the range of normal X-ray--emitting clusters.
Figure~\ref{fig:contour} shows the contour of the column density
distribution of gas and dark matter projected on a plane
in runs LR and MR.
The difference for the two runs is
already manifest,
especially in the gas distribution,
despite the fact that the only difference in the two runs
is the resolution.
In run MR the density profile is much more centrally concentrated.
Moreover, the cluster in run MR has two distinct subclumps,
which are not apparent in run LR.

\begin{figure}
\plotone{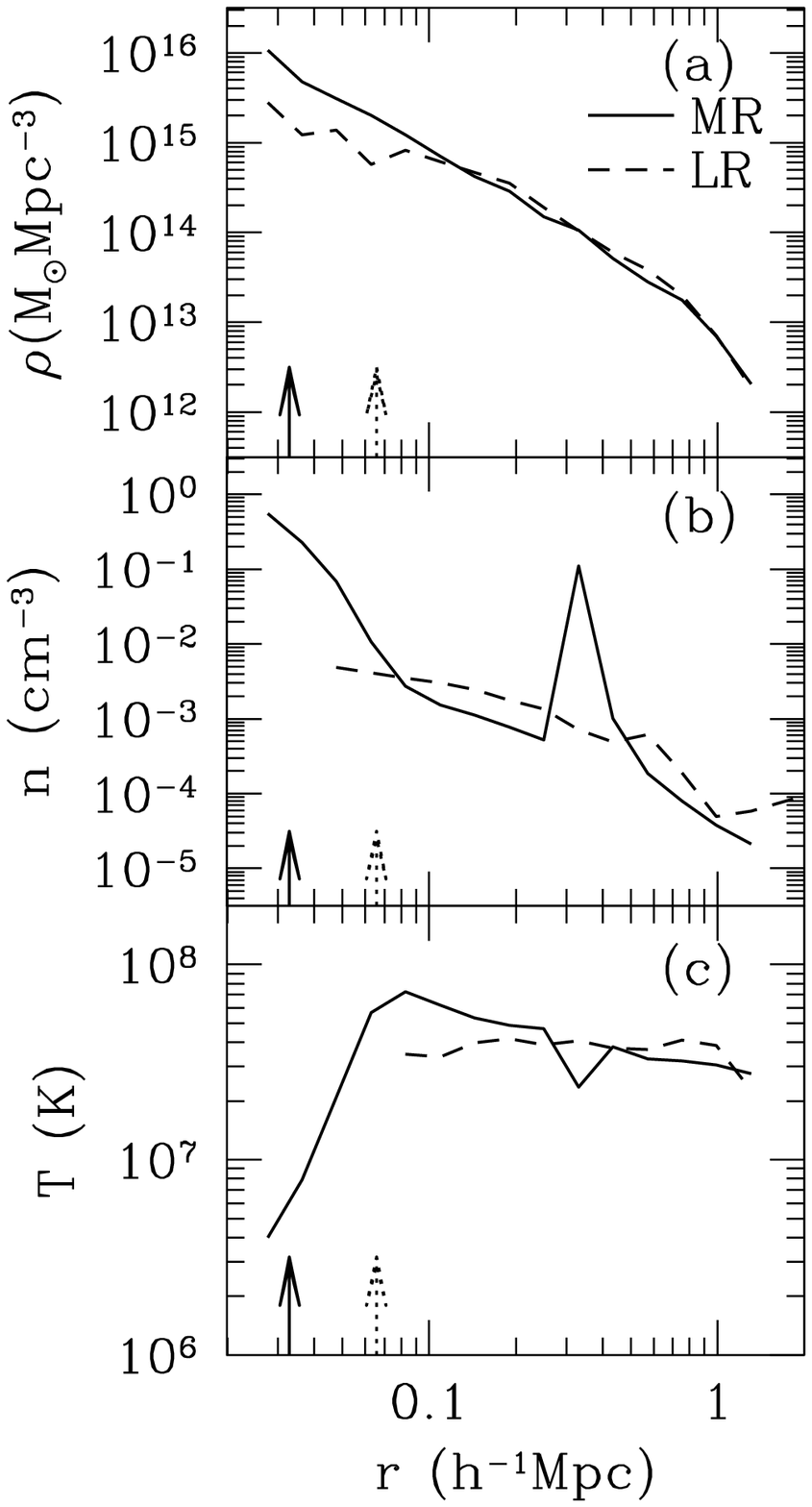}
\caption{Profiles of
(a) total mass density,
(b) gas density, and
(c) temperature
in the cluster in runs 
MR ({\it solid line}\/)
and 
LR ({\it dashed line}\/).
The abscissa is the physical radius.
The solid and dotted vertical arrows indicate
the gravitational softening length $\varepsilon$
for runs MR and LR, respectively.
\label{fig:profile}}
\end{figure}

The difference is also clear in Figure~\ref{fig:profile},
which shows total mass density $\rho$, gas density $n$,
and temperature $T$,
averaged over a spherical shell,
as functions of the radius.
In run LR, the logarithmic slope of the total mass density profile
gradually changes from $\sim -3$ in the outer part
to $\sim -1$ in the inner part,
i.e., the simulation fits
the NFW profile (eq.~[\ref{eqn:nfw}]).

The gas density profile in run LR
does not fit the NFW profile but
is well fitted by the 
observationally motivated
$\beta$ profile:
\begin{equation}
   n(r) = n_0 \biggl[
                 1 +
                 \Bigl(
                    \frac{r}{r_c}
                 \Bigr)^2
              \biggr]^{-3\beta/2},
\end{equation}
with
$n_0 = 5.4\times10^{-3}\,{\rm cm^{-3}}$,
$r_c = 140h^{-1}{\rm kpc}$,
and
$\beta = 0.75$.
The gas is nearly isothermal with $T=(3$--$4)\times10^7\,{\rm K}$
in run LR.
The X-ray luminosity in this run is 
$1.9\times10^{44}{\rm ergs\,s^{-1}}$
and is in the normal range observed.
This run has resolution and other properties typical of 
high-resolution numerical simulations of hydrodynamical cosmology.
If we had stopped our work at this point, we would have concluded that 
there was a good correspondence between observation and theory.

In contrast, in the higher resolution run MR
the total mass density profile 
shows a steeper rise toward the center
than in run LR.
The gas density profile also shows a steep rise toward the center
and cannot be fitted at all by the $\beta$ profile.
The temperature in run MR significantly drops toward the center.
There is a sharp peak
in the gas density
profile and drop in the temperature profile at $r\sim0.3h^{-1}{\rm Mpc}$;
this corresponds to the smaller subclump seen in
Figure~\ref{fig:contour}.
These results in run MR can naturally be interpreted as 
cooling of gas having decreased the temperature and increased the gas
density in the central part of the subclumps.

No observed clusters resemble the simulated cluster in run MR.
The X-ray intensity is smooth in the central region
in roughly half of the observed clusters,
which implies that the gas density is nearly constant.
The remainder of the clusters do show a peak
in X-ray emission, which is claimed to be a feature of a cooling
flow (\cite{fab91}).
However, the inferred gas density at the center of 
``cooling-flow'' clusters is much smaller than in our simulated cluster.
There is also a discrepancy between the simulation and the observations
in the total X-ray luminosity.
The bolometric luminosity $L_{\rm bol}$ of the cluster in run MR
is $7.4\times10^{45}\,{\rm ergs\,s^{-1}}$, 
far too high for a cluster with
temperature $\sim 3\,{\rm keV}$;
typically $L_{\rm bol} \sim 10^{44}\,{\rm ergs\,s^{-1}}$
for observed clusters with this gas temperature,
and none of these have bolometric luminosity exceeding 
$10^{45}\,{\rm ergs\,s^{-1}}$ (\cite{dav93}).

We have seen that
as a result of increasing the resolution
the simulated cluster has moved farther from the observed ones.
This is probably because cooling becomes too efficient.
The origin of this discrepancy is discussed in \S~\ref{sect:discussion}.

In general, the other 11 clusters in the higher resolution run have 
steep gas density profiles in the central part, similar to
the richest cluster.  
The central density of six of them is greater than
$10^{-1} {\rm cm^{-3}}$, 
and two of them have bolometric luminosity exceeding
$10^{45} {\rm ergs\,s^{-1}}$ 
although they have velocity dispersion of only
$500$--$600\,{\rm km\,s^{-1}}$, considerably smaller than the richest one.
Thus the properties of the richest cluster discussed above
are not to be attributed to a particular initial condition.

\subsection{Evolution}
\label{subsect:evolution}
\begin{figure}
\plotone{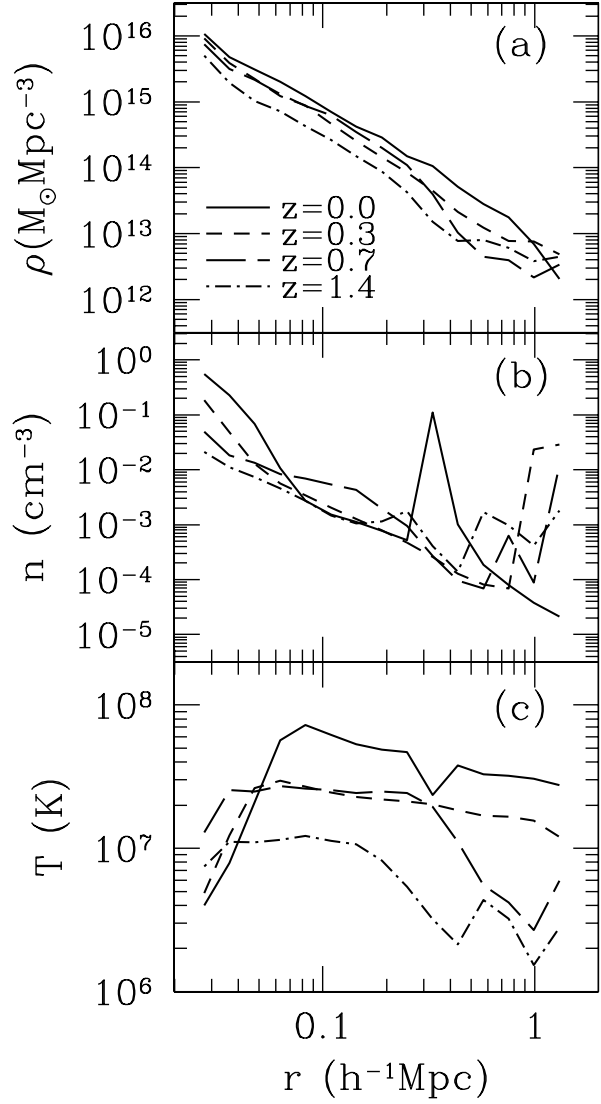}
\caption{Profiles of
(a) total mass density,
(b) gas density, and
(c) temperature
in the cluster in run MR
at 
$z=1.4$ ({\it dot-dashed line\/}),
$z=0.7$ ({\it long-dashed line\/}),
$z=0.3$ ({\it short-dashed line\/})
and
$z=0$ ({\it solid line\/}).
The abscissa is the physical radius.
\label{fig:evolution}}
\end{figure}

Next we examine
the time evolution of the cluster in run MR.
Figure~\ref{fig:evolution} shows the total mass density,
gas density, and temperature profiles at various redshifts.
The temperature increases at any radius
from $z=1.4$ to $z=0.7$, but
at $z \sim 0.7$ the temperature in the central part
begins to decrease, and it keeps
falling up to $z=0$.
The gas density in the central part increases rapidly
from $z\sim0.7$ to $z=0$.
We also notice that the smaller subclump is approaching the main
subclump.
These results support our view in the previous subsection
that cooling has resulted in a rise in the gas density profile 
toward the center.
On the other hand, the total mass density profile in the central part
is already quite steep at $z=1.4$.
This indicates that the steeper rise in run MR than in run LR 
is a consequence not of the effectiveness of cooling
but of the improved accuracy in following the evolution
of the dark matter.
As some authors have pointed out (\cite{fuk97}; \cite{moo98}),
the ``universal'' density profile of dark halos claimed by
Navarro, Frenk, and White (1996, 1997, eq.~[1]) 
is still a controversial issue 
and may depend on numerical resolution.
Clearly, our higher resolution run allowed the central density
to reach higher levels than the lower resolution run even in the
absence of cooling.
But then, in run MR the density reached a level where the cooling time
became short enough for the classical isobaric radiative cooling
instability to begin.
Allowance for metal-line cooling would have increased the radiative
losses by about a factor of 2 and would have accelerated
the trends noted.

We find that the cooling time in run LR is greater than the
Hubble time up to the resolution limit.
Therefore the main reason for the inefficiency of cooling in run
LR is simply that the achieved density is not high enough,
although two-body heating (\cite{ste97}) 
may have played an additional role in
preventing the gas from cooling.

\subsection{Timescales}
\label{subsect:timescale}
Now we investigate various timescales in the simulation
both as a function of radius and as a function of time.
The purpose is to confirm that
no obvious numerical
artifacts are invalidating the results of the
present simulations and that 
cooling is indeed what has caused the gas density profile to rise steeply
toward the center.

The timescales we are going to discuss are the following:
\begin{itemize}
\item Dynamical time:
\begin{equation}
   t_{\rm dyn} \equiv \frac{1}{(G\rho)^{1/2}}.
\end{equation}
\item Relaxation time due to artificial two-body effects:
\begin{equation}
   t_{\rm relax} 
      \equiv 0.34 \frac{\sigma^3}{G^2m_d\rho(\ln\Lambda)_g}
\end{equation}
(\cite{bin87}),
where $\sigma$ is the one-dimensional velocity dispersion
and $(\ln\Lambda)_g$ is the Coulomb logarithm associated with
gravitational interaction, which we estimate by
\begin{equation}
   (\ln\Lambda)_g = \ln(r/\varepsilon).
\end{equation}
\item Cooling time:
\begin{eqnarray}
   t_{\rm cool} 
      & \equiv & \frac{5}{2}
           \frac{nkT}{\Lambda_{\rm cool}} \nonumber \\
      & = & 2.1\times10^4 n^{-1}T^{1/2}\,{\rm yr},
\end{eqnarray}
where $k$ is Boltzmann's constant and
$n$ and $T$ in the second line are in cgs units.
\item Hubble time ($3/2$ times the age of the universe):
\begin{equation}
   t_H = H_0^{-1}(1+z)^{-3/2}.
\end{equation}
\end{itemize}

\begin{figure}
\plotone{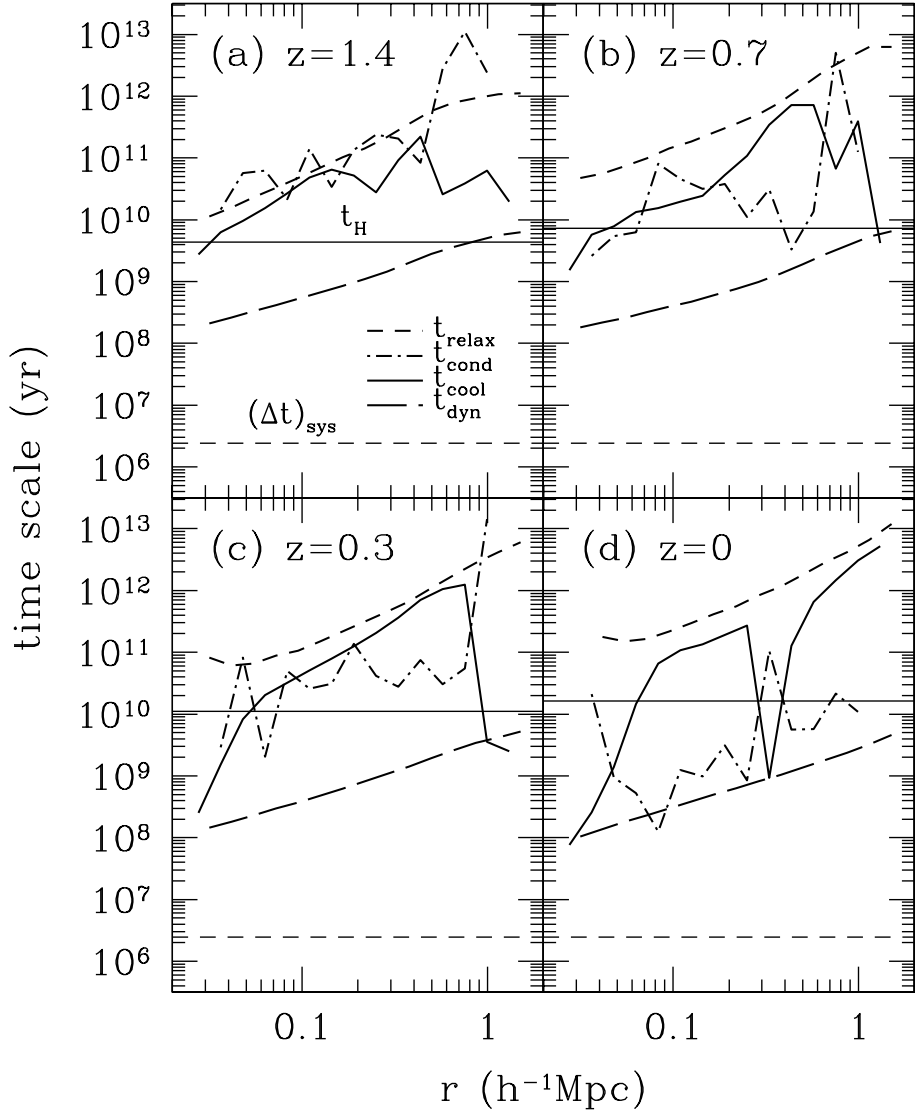}
\caption{
Dynamical time ({\it long-dashed line\/}),
two-body relaxation time ({\it short-dashed line\/}),
cooling time ({\it solid line\/})
and
conduction time ({\it dot-dashed line\/})
in the cluster in run MR
at
(a) $z = 1.4$,
(b) $z = 0.7$,
(c) $z = 0.3$,
and
(d) $z = 0$.
The horizontal thin solid line is the Hubble time, and
the horizontal thin dashed line is the system time step 
$(\Delta t)_{\rm sys}$.
The abscissa is the physical radius.
\label{fig:timescale}}
\end{figure}

Figure~\ref{fig:timescale} shows these timescales as functions of
radius at various redshifts.
Also shown is the heat conduction time $t_{\rm cond}$,
which will be considered in \S~\ref{sect:discussion}.  From 
Figure~\ref{fig:timescale} several points become clear:

\noindent
(i) All of the relevant timescales are always larger than
$(\Delta t)_{\rm sys}$ in any part of the cluster.
This assures that the time step adopted in our simulation
is short enough to accurately follow the evolution,
even where cooling is taking place.

\noindent
(ii) The relaxation time $t_{\rm relax}$ is always much greater than
the Hubble time $t_H$.
This means that the mass resolution is high enough that
the MR simulation is not being affected by artificial
two-body relaxation.

\noindent
(iii) The cooling time $t_{\rm cool}$ becomes shorter than $t_H$
at $z\sim0.7$ both at the center of the main clump
and at the center of the smaller subclump.
This implies that cooling is certainly affecting the thermal
and dynamical evolution of gas.

%
%
\section{Discussion}
\label{sect:discussion}

We have seen that the high-resolution
simulated clusters do not develop a core
in the gas distribution as is observed.
Rather, the gas density increases steeply toward the center.
In this section we discuss what is responsible for the 
discrepancy between the simulated and observed clusters.

To begin with, let us consider numerical resolution.
In fact it is uncertain whether we have yet achieved
sufficient mass and spatial resolution to accurately
follow the evolution of the gas density profile
with bremsstrahlung cooling included.
However, even if our resolution is insufficient, this is probably
not the cause of the discrepancy.
If we had better resolution, we would be able to follow
density inhomogeneities on small scales more accurately.
Since the cooling rate is proportional to the square of the gas
density, this would have resulted in more efficient cooling, and hence
an even steeper gas density profile toward the center.

%
In the remainder of the paper we discuss the effects of
(1) changing the cosmological parameters, 
(2) heat conduction,
and 
(3) heating of gas due to supernovae (SNe).

\subsection{Cosmological parameters}
\label{subsect:cosmparam}

There are now a variety of observations that constrain
the cosmological parameters, such as $h$, $\Omega_0$, $\Lambda_0$,
$\Omega_b$, $\sigma_8$, and $\Omega_\nu$,
where $\Omega_\nu$ is 
the density of an additional hot dark matter 
component in units of the critical density.
It seems reasonable to expect that changing these parameters
would not largely alter the cluster
density profile
if the parameters were constrained to satisfy observational requirements.

Since hot dark matter does not cluster on small scales,
adding a hot dark matter component will tend to make the central density
profile shallower.
This may push the simulation result closer to those of the observations.
However, too much of a hot dark matter component results in failure
to form objects at a high enough redshift,
and at the level of currently considered viable models,
$\Omega_\nu = 0.2$ (\cite{kly95}), the effects on cluster properties
are not large.

\subsection{Heat conduction}
\label{subsect:cond}

If heat conduction takes place in a timescale shorter than cooling time,
it will tend to
stabilize the gas, perhaps resulting in nearly isothermal gas
with a smooth density profile near the center.
We estimate the heat conduction time $t_{\rm cond}$ 
in the simulated cluster 
by
\begin{equation}
   t_{\rm cond} 
   \equiv \frac{5}{2}
     \frac{nkT}{|\nabla\cdot(\kappa\nabla T)|}.
\end{equation}
Here
\begin{equation}
   \kappa = 1.8 \times 10^{-5} T^{5/2} {(\ln\Lambda)_e}^{-1}
   \qquad {\rm ergs\,s^{-1}cm^{-1}K^{-1}}
\end{equation}
is the heat conductivity, where the Coulomb logarithm
$(\ln\Lambda)_e$, associated with electrostatic interactions
between electrons, is
\begin{equation}
   (\ln\Lambda)_e = 37.8 + \ln \biggl[ 
                           \Bigl(\frac{T}{10^8{\rm K}} \Bigr)
                           \Bigl(
                              \frac{n_e}{10^{-3}{\rm cm^{-3}}}
                           \Bigr)^{-1/2}
                           \biggr],
\end{equation}
with $n_e$ being the electron number density
(\cite{sar86}).

In Figure \ref{fig:timescale} we plot $t_{\rm cond}$ as a function
of radius.
Just outside the radius where $t_{\rm cool}$ begins to drop below
$t_H$, $t_{\rm cond}$ becomes smaller than both $t_{\rm cool}$
and $t_H$.
This implies that heat conduction may be efficient enough to prevent
the inner gas from cooling.

Thus heat conduction may have played a crucial role in the formation
of the core in spatial distribution of gas in clusters.
However, observations suggest that cluster gas is associated with
a weak magnetic field, which is tangled on a scale 
$\sim 10{\,\rm kpc}$.
If this is the case, the heat conductivity may be suppressed by
more than 2 orders of magnitude
(\cite{kro83}; \cite{cha98}), 
and conduction may not be able to
prevent gas from cooling.
At present, both the physical state of the magnetic field in clusters
(if existent) and the extent to which it can suppress conduction
remain open questions.

\subsection{SN heating}
\label{subsect:snheat}

Heat input to cluster gas
from supernova (SN) explosions occurring in cluster galaxies is
another process that may be responsible for the discrepancy between
the simulation and observations.
SN heating will raise the entropy and 
may lower the density and raise the
temperature of the gas so efficiently as to prevent it from cooling.
If this is the case, then the gas distribution may develop a core
similar to the observed ones.

\begin{figure}
\plotone{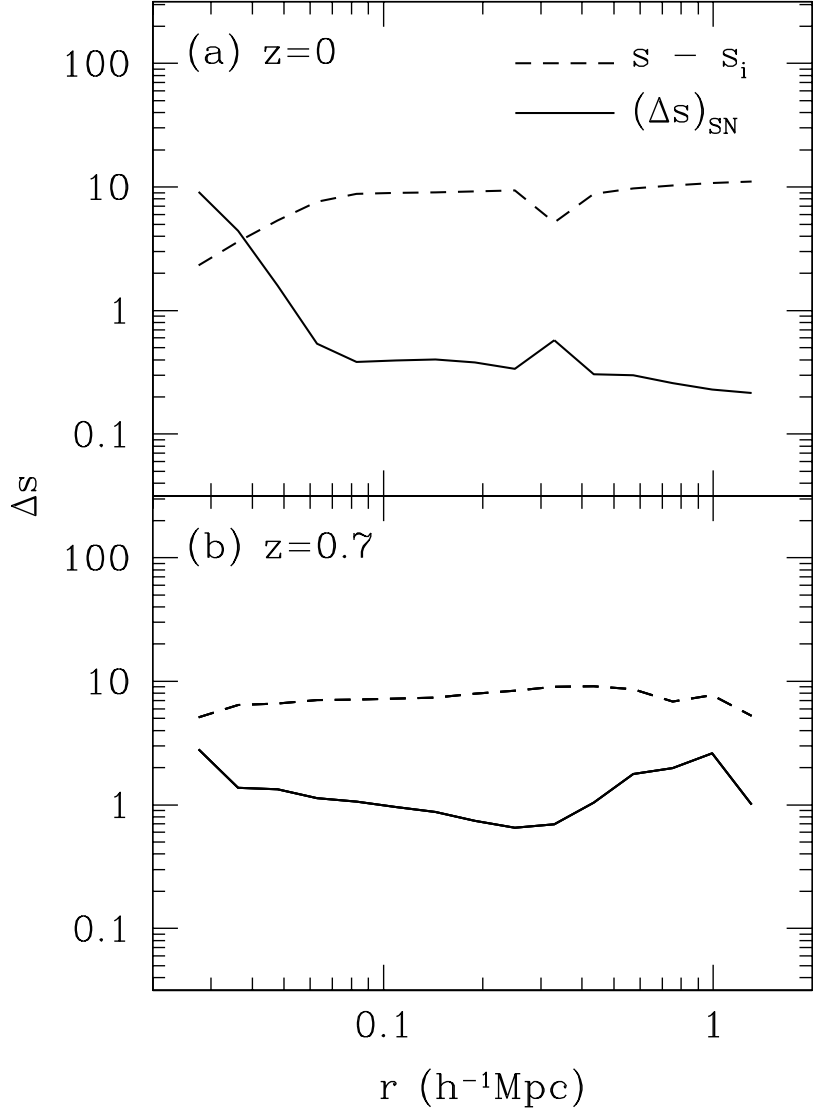}
\caption{
Entropy increase $(\Delta s)_{\rm SN}$ due to supernovae 
({\it solid line}),
and entropy change $s - s_i$ in the simulation
({\it dashed line}).
(a) $z=0$.
(b) $z=0.7$.
\label{fig:snheat}}
\end{figure}

To assess this effect, we estimate entropy increase due to
SN heating, using observed iron abundance.
Heat that has been ejected from SNe per unit gas mass
can be estimated as
\begin{equation}
   \Delta q = 2.0\times10^{-3} \chi 
                 \biggl[
                    f_{\rm I}\frac{E_{\rm I}}{M_{\rm Fe, I}}
                    + ( 1 - f_{\rm I} )
                      \frac{E_{\rm II}}{M_{\rm Fe, II}}
                 \biggr] ,
\end{equation}
where $\chi$ is the iron abundance in units of solar abundance,
$E_{\rm I}$ and $M_{\rm Fe, I}$ are the explosion energy and
ejected iron mass per Type Ia SN,
$E_{\rm II}$ and $M_{\rm Fe, II}$ are the same for Type II SN,
and
$f_{\rm I}$ is the fraction of iron that has been produced by 
Type I SNe.
We adopt
\begin{eqnarray}
   E_{\rm I} & = & 1.3\times10^{51}\,{\rm ergs},\\
   M_{\rm Fe, I} & = & 0.744\,M_{\odot},\\
   E_{\rm II} & = & 1\times10^{51}\,{\rm ergs},
\end{eqnarray}
and
\begin{equation}
   M_{\rm Fe, II}  =  0.038\,M_{\odot}
\end{equation}
(\cite{nom84}; \cite{thi93}).
The value of $M_{\rm Fe, II}$ 
is an average over the progenitor mass
weighted with the Salpeter
initial mass function.
The fraction $f_{\rm I}$ can be estimated from abundance ratios,
but there is still considerable uncertainty
(\cite{gib97}).
We adopt $f_{\rm I} = 0.5$.
We take the AWM7 cluster as an example (\cite{eza97})
and fit the observed iron abundance as a function of radius $r$ to
the following profile:
\begin{equation}
   \chi = 0.58 \biggl[
                  1 + \Bigl(\frac{r}{0.058h^{-1}{\rm Mpc}}\Bigr)^2
               \biggr]^{-0.3}.
\end{equation}
If the SN rate is nearly constant up to the present epoch,
we can approximate the entropy increase due to SNe,
$(\Delta s)_{\rm SN}$, fairly well by $\Delta q/T$.
Figure \ref{fig:snheat}$a$ shows $(\Delta s)_{\rm SN}$
as a function of radius $r$.
We express the entropy in units of $3k/(2\mu m_p)$,
where $\mu$ is the mean molecular weight and
$m_p$ is the proton mass.
We also plot $s-s_i$, where $s_i$ and $s$ are the entropy
at the initial and present epochs in the simulation.
In the innermost region of the cluster, $(\Delta s)_{\rm SN}$
exceeds $s-s_i$.
This implies that heating due to SNe can significantly alter the thermal
evolution of the gas in the central region.

If the SN rate was much larger at $z \gtrsim 1$ than at $z \sim 0$, 
as is suggested by apparent lack of evolution 
in the iron abundance in clusters toward higher redshifts
(\cite{mus97}),
then the above estimate of
$(\Delta s)_{\rm SN}$ is a poor approximation in the central region
because the temperature has significantly decreased toward the present
epoch.
In this case, it is better to assess the effect at a higher redshift.
Figure \ref{fig:snheat}$b$ shows $(\Delta s)_{\rm SN}$ and
$s-s_i$ at $z=0.7$.
This time $(\Delta s)_{\rm SN}$ is smaller than $s-s_i$ everywhere.
This indicates that SN heating should have little effect 
in cases in which SN activities
in the cluster continued at a nearly constant rate up to
$z\sim 0.7$ and have essentially stopped after that.
Finally, we can consider the case where most of the entropy
was injected within the first $\sim 10^9{\rm yr}.$
Then the effect should be very significant,
because the gas temperature was relatively low:
we find $n\sim1\times 10^{-3}{\rm cm^{-3}}$ and 
$T\sim 5\times10^5\,{\rm K}$
at the density peak of a protocluster at $z=5$, which will end up in
the central region of the cluster at $z=0$.
Hence $(\Delta s)_{\rm SN}$ is estimated as $\sim 80$
and far exceeds $s-s_i\sim 4$.

Thus SN heating can be an important factor in the formation of
the core in the gas distribution, but the effect is strongly
dependent on, among other things, 
the detailed star formation history inside
the cluster.

%
%
\section{Conclusions}
\label{sect:conclusion}

Finally let us briefly summarize our conclusions.
\begin{enumerate}
\item Our multiresolution simulation has followed 
with reasonable accuracy the evolution of
gas and dark matter in a typical cluster
under the influence of bremsstrahlung cooling 
as well as gravity and hydrodynamics.
\item The high resolution simulation resulted in a gas density profile
steeply rising toward the center,
with consequent very high X-ray luminosity; 
however, these properties are not observed.
\item Heat conduction and SN heating are among the
processes that may account for the discrepancy.
Had we allowed for their likely importance in the real world
we might have been able to
recover the observed
gas density profile.
In a future work we will examine their effects by directly
incorporating them into the simulation.
\end{enumerate}

\acknowledgments
T.~S. acknowledges support from Research
Fellowships of the Japan Society for the Promotion of Science.
J.~P.~O. was supported by NASA grant NAG~5-2759
and NSF grants AST~93-18185, ACI~96-19019 (PACI Subaward No.~766), 
and AST~94-24416.
Numerical computation was carried out on VPP/300 and VX/4R
at the Astronomical Data Analysis Center of
the National Astronomical Observatory in Japan,
as well as on VX/4R at RESCEU
(Research Center for the Early Universe, the University of Tokyo).
\end{document}